\documentclass[preprint,review,11pt]{elsarticle}
\RequirePackage[left=2.5cm,right=2.0cm,top=3.0cm,bottom=3.0cm,a4paper]{geometry}
 
\RequirePackage{bm}

\usepackage{amsthm} 
\usepackage{amssymb,amsmath,mathrsfs}
\usepackage{algpseudocode}
\usepackage{algorithmicx,algorithm}
\usepackage{subfigure}
\usepackage{setspace}

\usepackage{graphicx}
\usepackage{xcolor}
\usepackage[colorlinks,linkcolor=red,anchorcolor=blue,citecolor=green,CJKbookmarks=True]{hyperref}
\usepackage{booktabs}
\usepackage{lineno}
\renewcommand{\arraystretch}{1.5}

\journal{}

\begin{document}

\begin{frontmatter}

\title{High-order Discontinuous Galerkin solver based on Jacobi polynomial expansion for compressible flows on unstructured meshes}

\author[label1]{Yu-Xiang Peng}
\ead{pengyx55@mail.sysu.edu.cn}
\author[label1]{Biao Wang}
\ead{wangb228@mail2.sysu.edu.cn}
\author[label1]{Peng-Nan Sun\corref{cor1}}
\ead{sunpn@mail.sysu.edu.cn}
\author[label2]{A-Man Zhang}
\ead{zhangaman@hrbeu.edu.cn}

\address[label1]{School of Ocean Engineering and Technology, Sun Yat-sen University, Zhuhai 519082, China}
\address[label2]{College of Shipbuilding Engineering, Harbin Engineering University, Harbin 150001, China}
\cortext[cor1]{Corresponding author: sunpn@mail.sysu.edu.cn}

\begin{abstract}
Based on the Jacobi polynomial expansion, an arbitrary high-order Discontinuous Galerkin solver for compressible flows on unstructured meshes is proposed in the present work. First, we construct orthogonal polynomials for 2D and 3D isoparametric elements using the 1D Jacobi polynomials. We perform modal expansions of the state variables using the orthogonal polynomials, enabling arbitrary high-order spatial discretization of these variables. Subsequently, the discrete governing equations are derived by considering the orthogonality of the Euler equations' residuals and the test functions. On this basis, we develop a high-order Discontinuous Galerkin solver that supports various element types, including triangles, quadrilaterals, tetrahedra, hexahedra, etc. An improved shock-capturing scheme has been adopted to capture shock discontinuities within the flow field. The variable's gradients at the discontinuous elements are reconstructed by its adjacent elements, and the slope limiter is applied to modify the state variables, smoothing the state variables and enhancing the robustness of the solver. The convergence rates of solvers of different orders have been verified by a benchmark case, and the CPU costs are given to prove that high-precision algorithms have higher computational efficiency under the same error level. Finally, several two- and three-dimensional compressible fluid dynamics problems are studied, compared with literature and experimental results, the effectiveness and accuracy of the solver were verified.
\end{abstract}


\end{frontmatter}


\section{Introduction}
\label{sec-introduction}
Computational fluid dynamics (CFD) methods have been widely used in engineering and have become increaselingly important in recent years \cite{tu2023computational}. The widely used numerical method in the engineering application is finite volume method(FVM), the traditional FVM solver typically has first-order or second-order accuracy. With the increasing demand for flow-field analysis accuracy, many high-order methods have been proposed \cite{wang2013high}. For instance, the WENO reconstruction scheme in finite volume and finite difference framework \cite{wang2021new,chen2022physical,zhang2024finite}, as well as spectral element method \cite{cantwell2015nektar++}, discontinuous Galerkin method \cite{cockburn2012discontinuous,liu2022essentially}, etc.

Among these high-order methods mentioned above, the WENO reconstruction is the most widely used since it can be directly extended in the original finite volume code. Unlike the WENO reconstruction, which only performs high-order reconstruction on the numerical flux at the element boundaries, spectral element method and discontinuous Galerkin method perform high-order expansion both inside and at the boundary of the element, the influence of the non-uniformity distribution of variables inside the element is considered \cite{di2011mathematical}. The spectral element method usually assumes that the physical quantities between elements are continuous, while discontinuous Galerkin does not limited by this assumption, making it more suitable for dealing with strong compressible discontinuity problems. It has been widely applied to compressible aerodynamics \cite{silveira2015higher}, transitional turbulent flows \cite{fernandez2017hybridized}, and multiphase flows \cite{owkes2013discontinuous,ge2020investigation,zhang2023analysis}.

The discontinuous finite element method (DG-FEM) was first proposed by Reed and Hill in 1973 and applied to solve the neutron transport equation \cite{reed1973triangular}. Cockburn and Shu extended the discontinuous Galerkin method to solve the nonlinear conservation laws, attracting widespread attention \cite{cockburn1989tvb,cockburn1998runge}. Subsequently, researchers have developed various discontinuous Galerkin methods for solving different equations, such as local discontinuous Galerkin \cite{yan2002local}, reconstructed discontinuous Galerkin \cite{luo2010reconstructed} and direct discontinuous Galerkin \cite{danis2022new}. For a more detailed introduction to discontinuous Galerkin, reader may refer to the review by Cockburn et al. \cite{shu2014discontinuous}.

The accuracy and convergence of the discontinuous Galerkin method (DGM) have been proven mathematically. However, the specific numerical implementation of arbitrary high-order DGM for unstructured grids is not detailed enough. Moreover, in most research papers related to DGM, usually, only the quadratic polynomials are constructed, and changing the order of the solver requires reconstructing new basis functions. Which is very inconvenient for implementing arbitrarily high-order solvers and is not aesthetically pleasing mathematically.

In this study, similar to the spectral element method \cite{karniadakis2005spectral}, we use orthogonal polynomials to construct high-order spatial expansions of variables. Specifically, the arbitrary high-order orthogonal polynomials in finite element isoparametric elements are derived based on one-dimensional Jacobi polynomials. Using orthogonal polynomials as basis functions, expand the state variables within the elements to achieve high-order spatial discretization. This scheme can easily construct arbitrary high-order solver and maintain theoretical consistency with the Gauss quadrature. On this basis, we introduce the variable normal projection algorithm for solving numerical flux at the boundary of two-dimensional and three-dimensional unstructured grids. The trouble-cell detection scheme was adopted to mark the discontinuous elements \cite{fu2017new}, and a new reconstruction algorithm for the gradient at discontinuous elements was proposed. The discontinuous region was smoothed by limiting the variables' gradient, and the computational stability was increased.

This paper is organized as follows. In Section 2, we first review the governing equations of inviscid compressible flow and construct orthogonal polynomials on isoparametric elements using 1D Jacobi polynomials. We then discretize the governing equations by multiplying the basis functions and carefully derive the numerical flux at the boundaries of nonstructured elements. In Section 3, we introduce the interaction scheme for both the interior and boundary of the elements, along with the implementation of shock detection and limiters. Section 4 verifies the efficiency and accuracy of the proposed methods. Several numerical examples are studied in Section 5 to validate the effectiveness and accuracy of the present solver. Finally, a summary of this paper is provided in Section 6.

\section{Basic theory of Discontinuous Galerkin Method (DGM)}
\subsection{Governing equations for the compressible fluid dynamics}
For the compressible and inviscid fluid dynamics, the fluid flow is controlled by the so-called Euler equations \cite{anderson1995computational}, which can be expressed shortly with the material derivative notation as follows:
\begin{equation}
\left\{
\begin{aligned}
\frac{D\rho}{Dt} &=-\rho\nabla\cdot{\bm{u}} \\
\frac{D\bm{u}}{Dt} &=-\frac{1}{\rho}\nabla p \\
\frac{De}{Dt} &=-\frac{p}{\rho}\nabla\cdot\bm{u}
\end{aligned}
\right.
\end{equation}

\noindent The above equations are the mass, momentum and energy conservation equations, respectively. Where $t$ is the time and '$D$' denotes the material derivative. $\rho$, $\bm{u}$, and $p$ are the primitive variables, which denote the density, velocity, and pressure of the fluid, respectively. $e$ is the specific internal energy (means the internal energy per unit mass). These equations can be regarded as the simplification of the Navier-Stokes equation. And the system is closed by the equation of state (EOS), in the present work the stiffened equation of state is adopted:
\begin{equation}
p=(\gamma-1)\rho e+\gamma B
\end{equation}

\noindent where $\gamma$ and $B$ are the material constants, for the monoatomic gas $\gamma=1.4$ and $B=0$. And the speed of sound $c$ is calculated by $c=\sqrt{{\gamma(p+B)}/{\rho}}$. For the computational fluid dynamics, the above governing equation is rewritten in the conservation form, which is expressed as (three-dimensional):
\begin{equation}
\label{eq-euler-equation-conservation-form-1d}
\left\{
\begin{aligned}
\frac{\partial \rho}{\partial t}+\frac{\partial \rho u}{\partial x}+\frac{\partial \rho v}{\partial y}+\frac{\partial \rho w}{\partial z}=0\\
\frac{\partial \rho u}{\partial t}+\frac{\partial \rho u^{2}+p}{\partial x}+\frac{\partial \rho uv}{\partial y}+\frac{\partial \rho uw}{\partial z}=0\\
\frac{\partial \rho v}{\partial t}+\frac{\partial \rho uv}{\partial x}+\frac{\partial \rho v^{2}+p}{\partial y}+\frac{\partial \rho vw}{\partial z}=0\\
\frac{\partial \rho w}{\partial t}+\frac{\partial \rho uw}{\partial x}+\frac{\partial \rho vw}{\partial y}+\frac{\partial \rho w^{2}+p}{\partial z}=0\\
\frac{\partial E}{\partial t}+\frac{\partial u(E+p)}{\partial x}+\frac{\partial v(E+p)}{\partial y}+\frac{\partial w(E+p)}{\partial z}=0
\end{aligned}
\right.
\end{equation}

\noindent where $E$ is the total energy per unit volume, which equates to the sum of the potential energy generated by the internal pressure and the kinetic energy of the gas, we have:
\begin{equation}
E=\rho(e+\frac{1}{2}|\bm{u}|^2)=\frac{p}{\gamma-1}+\frac{1}{2}\rho (u^{2}+v^{2}+w^{2})
\end{equation}

\noindent To construct the relative generalized numerical discretization of a class of nonlinear conservation laws, the above Euler equations are expressed as the following vector form:
\begin{equation}
\frac{\partial\bm{Q}}{\partial t}+\frac{\partial\bm{F}(\bm{Q})}{\partial x}+\frac{\partial\bm{G}(\bm{Q})}{\partial y}+\frac{\partial\bm{H}(\bm{Q})}{\partial z}=0
\end{equation}

\noindent in which the vector $\bm{Q}$ is the conservative state vector, $\bm{F}$, $\bm{G}$ and $\bm{H}$ are the fluid nonlinear fluxes, respectively. We have:
\begin{equation}
\renewcommand{\arraystretch}{1.0}
\bm{Q}=\left(
\begin{array}{c}
{\rho }\\{\rho u}\\{\rho v}\\{\rho w}\\{E}
\end{array}\right),\,\,
\bm{F}=\left(
\begin{array}{c}
{\rho u}\\{\rho u^{2}+p}\\{\rho uv}\\{\rho uw}\\{u(E+p)}
\end{array}\right),\,\,
\bm{G}=\left(
\begin{array}{c}
{\rho v}\\{\rho uv}\\{\rho v^{2}+p}\\{\rho vw}\\{v(E+p)}
\end{array}\right), \,\,
\bm{H}=\left(
\begin{array}{c}
{\rho w}\\{\rho uw}\\{\rho vw}\\{\rho w^{2}+p}\\{w(E+p)}
\end{array}\right)
\end{equation}

\noindent Finally, the governing equation for the compressible fluid is obtained, and the system is closed by adding the equation of state. All the effort is to solve the fluid state vector by numerical methods, and the Discontinuous Galerkin Method is adopted in the present work.

\subsection{Modal expansion}
Firstly, the physical domain $\varOmega$ is approximated by the computational domain $\varOmega_h$ which is discretized into a series of elements $\mathcal{D}_k$, $k$ denotes the index of the element. The approximate solution $\bm{Q}^h$ is assumed to be a piecewise continuous function in the finite element space, which is to say $\bm{Q}^h$ is continuous within each element. Still, the discontinuity is allowed between two elements, and $\bm{Q}^h\in L^\infty(\varOmega)$. In the present work, the Discontinuous Galerkin Method (DGM) is adopted to solve the approximate state vector $\bm{Q}^h$. In each DGM element, we assume the approximate solution $\bm{Q}^h_k\in V_h(\mathcal{D}_k)$, where the local space $V_h(\mathcal{D}_k)={\rm{span}}\left\{\phi_n(\mathcal{D}_k)\right\}^{N_p}_{n=1}$ is the collection of polynomials, and $\phi_n(\mathcal{D}_k)$ denotes $n$-th polynomial basis. By using the modal expansion \cite{karniadakis2005spectral}, the approximate state vector can be expressed as follows:
\begin{equation}
\label{eq-modal-expandsion}
\bm{x}\in\mathcal{D}_k : \bm{Q}^h(\bm{x}(\xi,\eta,\zeta),t) =\sum^{N_p}_{n=1}\bm{Q}^k_n(t)\phi_n(\xi,\eta,\zeta)
\end{equation}

\noindent where $\bm{x}$ denotes the coordinate of the point in $\mathcal{D}_k$, and $t\in [0, T]$ denotes the time. $\bm{Q}^k_n(t)$ denotes the coefficient for the modal expansion, which is the local unknowns, and there are $N_p$ unknown state vectors to be solved on each element. $(\xi, \,\eta,\, \zeta)$ in the basis function denotes the natural coordinate of the element $\mathcal{D}_k$. Adopting the parametric transformation \cite{zienkiewicz2005finite}, the natural and Cartesian coordinates have the following relation:
\begin{equation}
\bm{x}=\sum_{i=1}^{N_v}\bm{x}_iN_i(\bm{\xi})
\end{equation}

\noindent where $N_v$ is the number of the element vertices, $\bm{x}_i$ is the coordinates of the $i$-th vertex, $N_i(\bm{\xi})$ denotes the shape function. For the parameter transformation of other isoparametric elements, such as triangle, quadrilateral and hexahedron elements, etc., readers may refer to \cite{cook2007concepts}. In the present work, the normalized Jacobi polynomial is adopted to construct the basis function, which is suitable for developing arbitrary high-order algorithms. The initial $n$-th order Jacobi polynomial \cite{szeg1939orthogonal} can be calculated by the Rodrigues formula:
\begin{equation}
J_{n}^{(\alpha,\beta)}(\xi) = \frac{(-1)^{n}}{2n\cdot n!}(1-\xi)^{-\alpha}(1+\xi)^{-\beta}\frac{d^{n}}{d\xi^{n}}\left[(1-\xi)^{n+\alpha}(1+\xi)^{n+\beta}\right]
\end{equation}

\noindent where $\alpha>-1,\beta >-1$ are two pre-defined parameters. The Jacobi polynomials satisfy the following orthogonal condition:
\begin{equation}
\label{eq-orthogonality-condition-jacobi-poly}
\int_{-1}^{1}\omega^{(\alpha,\beta)}(\xi)J_m^{(\alpha,\beta)}(\xi)J_n^{(\alpha,\beta)}(\xi)d\xi = \gamma_{n}^{(\alpha,\beta)}\delta_{mn}
\end{equation}

\noindent where $\omega^{(\alpha,\beta)}(\xi)=(1-\xi)^\alpha(1+\xi)^{\beta}$ is the weight function for the orthgal polynomials, and $\gamma^{\alpha,\beta}_n$ is constant
\begin{equation}
\gamma_{n}^{(\alpha,\beta)} =  \frac{2^{\alpha+\beta+1}\Gamma(n+\alpha+1)\Gamma(n+\beta+1)}{(2n+\alpha+\beta+1)\Gamma(n+1)\Gamma(n+\alpha+\beta+1)}
\end{equation}

\noindent In one dimension case, by using the normalized Jacobi polynomials, the orthogonal basis function can be constructed:
\begin{equation}
\label{eq-basis-function}
\phi_n(\xi)=\sqrt{\omega^{(\alpha,\beta)}(\xi)}P^{(\alpha,\beta)}_{n}(\xi)\, ,\, \text{in which} \,:\, P^{(\alpha,\beta)}_n(\xi)=\frac{J^{(\alpha,\beta)}_{n}(\xi)}{\sqrt{\gamma_n^{(\alpha,\beta)}}}
\end{equation}

\noindent where $P^{(\alpha,\beta)}_n(\xi)$ is the normalized Jacobi polynomial. And the derivatives of the Jacobi polynomial can be expressed as follows: 
\begin{equation}
\label{eq-derivative-jacobi-poly}
\frac{d}{d \xi}\left[P_{n}^{(\alpha, \beta)}(\xi)\right]=\sqrt{n(n+\alpha+\beta+1)} P_{n-1}^{(\alpha+1, \beta+1)}(\xi)
\end{equation}

\noindent Combine Eq. \eqref{eq-basis-function} and the orthogonality relation of the Jacobi polynomial Eq. \eqref{eq-orthogonality-condition-jacobi-poly}, the following orthogonality property of the basis function is obtained:
\begin{equation}
\int_{-1}^{1}\phi_{m}(\xi)\phi_{n}(\xi)d\xi=\delta_{mn}
\end{equation}

\noindent Considering that for the isoparametric element in one dimension, we have $\xi\in[-1,1]$. Thus, the above equation means that the basis function is orthogonal over the isoparametric element. In the present work, we set $\alpha=0$ and $\beta=0$ for simplicity. Therefore, $\omega^{(\alpha,\beta)}(\xi)=1$, and $\displaystyle\gamma^{(\alpha,\beta)}_n=2/({2n+1})$, the Jacobi polynomial is simplified to the Langende polynomial, $\phi_n(\xi)=P^{(0,0)}_n(\xi)=\sqrt{(2n+1)/2}J^{(0,0)}_n(\xi)$. Then the derivative of the basis function is:
\begin{equation}
\frac{d}{d\xi}\phi_{n}(\xi)=\sqrt{n(n+1)}J_{n-1}^{(1,1)}(\xi)
\end{equation}

\noindent In quadrilateral and hexahedral elements, the basis function can be directly constructed by the tensor product. We have:
\begin{align}
\text{For quadrilateral elements:}\,\,  \phi_n(\xi,\eta) &= P_i^{(0,0)}(\xi)P_j^{(0,0)}(\eta)  \\
\text{For hexahedron elements:}\,\,  \phi_n(\xi,\eta,\zeta) &=P_i^{(0,0)}(\xi)P_j^{(0,0)}(\eta)P_k^{(0,0)}(\zeta) 
\end{align}

\noindent where $0\le i,j,k\le N$, $N$ denotes the highest order of the normalized Jacobi polynomial, and $N$ determinates the convergence rate of the numerical algorithm. The above basis function is orthogonal on the isoparametric elements, and the isoparametric quadrilateral element is a square in 2D: $\mathcal{I}'=\left\{(\xi,\eta)|\xi,\eta\ge-1,\xi,\eta\le1\right\}$, the isoparametric hexahedron element is a cube in 3D: $\mathcal{I}'=\left\{(\xi,\eta,\zeta)|\xi,\eta,\zeta\ge-1,\xi,\eta,\zeta\le1\right\}$. The derivative of the basis function can be derived by the chain rule, and for more details, readers may refer to \cite{szeg1939orthogonal}. The construction of the orthogonal basis functions on triangles and tetrahedrons is slightly more complicated, the Proriol–Koornwinder–Dubiner (PKD) polynomials\cite{proriol1957famille,koornwinder1975two,dubiner1991spectral} which construct by Jacobi polynomial are adopted to be the basis function, which is expressed as follows:
\begin{align}
\text{For triangle element:} \,\,  \phi_n(\xi,\eta) &=2\sqrt{2}P^{(0,0)}_i(\xi)P^{(2i+1,0)}_{j}(\eta)(1-\eta)^i \\
\text{For wedge element:} \,\,  \phi_n(\xi,\eta,\zeta) &= 2\sqrt{2} P^{(0,0)}_i(\xi)P^{(2i+1,0)}_{j}(\eta)P^{(0,0)}_{k}(\zeta)(1-\eta)^i \\
\text{For tetrahedron element:} \,\,  \phi_n(\xi,\eta,\zeta) &=8 P^{(0,0)}_i(\xi)P^{(2i+1,0)}_{j}(\eta)P^{(2i+2j+2,0)}_{k}(\zeta)(1-\eta)^i(1-\zeta)^{i+j}
\end{align}

\noindent The above basis function is orthogonal on the isoparametric elements, and the isoparametric quadrilateral element is a square in 2D: $\mathcal{I}'=\left\{(\xi,\eta)|\xi,\eta\ge 0,\xi+\eta\le1\right\}$, the isoparametric hexahedron element is a cube in 3D: $\mathcal{I}'=\left\{(\xi,\eta,\zeta)|\xi,\eta,\zeta\ge 0,\xi+\eta+\zeta\le1\right\}$. Similarly, the derivative of the basis functions can be derived by the chain rule.

\subsection{The discretization of the governing equation}
The state vector of the fluid field in the $k$-th element can be expanded by Eq.\eqref{eq-modal-expandsion}. Considering the orthogonality property of the basis function, the value of the expansion coefficients $\bm{Q}_n(t)$ in the above equation can be calculated by:
\begin{equation}
\label{eq-modal-expandsion-coefficient}
\bm{Q}_n(t)=\int_{\mathcal{D}^k}\bm{Q}(\bm{x},t)\phi_n(\xi,\eta,\zeta)d\varOmega
\end{equation}

\noindent The above equation can be adopted to obtain the initial value of the expansion coefficients. Let $\bm{Q}^h$ be the approximate solution of the state variables, thus, the residual of the Euler equation is:
\begin{equation}
\mathcal{R}^h(\bm{x}, t):= \frac{\partial\bm{Q}^h}{\partial t}+\frac{\partial\bm{F}(\bm{Q}^h)}{\partial x}+\frac{\partial\bm{G}(\bm{Q}^h)}{\partial y}+\frac{\partial\bm{H}(\bm{Q}^h)}{\partial z} \quad \forall \bm{x}\in \mathcal{D}^k
\end{equation}

\noindent Since the residual is asked to be orthogonal to the test functions in the Galerkin method \cite{bangerth2013adaptive}, the basis functions are chosen to be the test functions. Therefore, we have:
\begin{equation}
\int_{\mathcal{D}^k}\mathcal{R}^h(\bm{x}, t)\phi_nd\varOmega= \int_{\mathcal{D}^k}\left(\frac{\partial\bm{Q}^h}{\partial t}+\frac{\partial\bm{F}(\bm{Q}^h)}{\partial x}+\frac{\partial\bm{G}(\bm{Q}^h)}{\partial y}+\frac{\partial\bm{H}(\bm{Q}^h)}{\partial z}\right)\phi_nd\varOmega=0
\end{equation}

\noindent Considering the modal expansion of the state vector Eq.\eqref{eq-modal-expandsion}, and using the chain rule, the above equation is rewritten as:
\begin{equation}
\begin{aligned}
\int_{\mathcal{D}^k}\frac{\partial \bm{Q}_m}{\partial t}\phi_m\phi_nd\varOmega &+\int_{\mathcal{D}^k}\left(\frac{\partial \bm{F}(\bm{Q}^h)\phi_n}{\partial x}+\frac{\partial \bm{G}(\bm{Q}^h)\phi_n}{\partial y}+\frac{\partial \bm{H}(\bm{Q}^h)\phi_n}{\partial z}\right)d\varOmega \\
&-\int_{\mathcal{D}^k}\left(\bm{F}(\bm{Q}^h)\frac{\partial \phi_n}{\partial x}+\bm{G}(\bm{Q}^h)\frac{\partial \phi_n}{\partial y}+\bm{H}(\bm{Q}^h)\frac{\partial \phi_n}{\partial z}\right)d\varOmega=0
\end{aligned}
\end{equation}

\noindent By using Gauss's theorem, we obtain the local statement:
\begin{equation}
\begin{aligned}
\int_{\mathcal{D}^k}\phi_m\phi_nd\varOmega\frac{\partial \bm{Q}_m}{\partial t} & +
\int_{\partial\mathcal{D}^k}[{n}_x\bm{F}(\bm{Q}^h)+{n}_y\bm{G}(\bm{Q}^h)+{n}_z\bm{H}(\bm{Q}^h)]\phi_nd\varGamma \\
&-\int_{\mathcal{D}^k}\left(\bm{F}(\bm{Q}^h)\frac{\partial \phi_n}{\partial x}+\bm{G}(\bm{Q}^h)\frac{\partial \phi_n}{\partial y}+\bm{H}(\bm{Q}^h)\frac{\partial \phi_n}{\partial z}\right)d\varOmega=0
\end{aligned}
\end{equation}

\noindent where $\bm{n}={n}_x\bm{e}_1+{n}_y\bm{e}_2+{n}_z\bm{e}_3$ is the unit outward normal at the element boundary. $\bm{F}(\bm{Q}^h), \bm{G}(\bm{Q}^h), \bm{H}(\bm{Q}^h)$ in the third term at the left hand side is the physic flux in the element. In DGM, since the boundary $\partial\mathcal{D}^k$ might be shared by two elements, which means the physic flux is multi-defined. To consider the interaction between elements, the flux ${n}_x\bm{F}(\bm{Q}^h)+{n}_y\bm{G}(\bm{Q}^h)+{n}_z\bm{H}(\bm{Q}^h)$ in the second term is replaced by the numerical flux $\left[{n}_x\bm{F}(\bm{Q}^h)+{n}_y\bm{G}(\bm{Q}^h)+{n}_z\bm{H}(\bm{Q}^h)\right]^*$, which can be calculated by the Riemann Solver and is discussed in the following section. Thus, we have:
\begin{equation}
\label{eq-DGM-governing-equation}
\begin{aligned}
\int_{\mathcal{D}^k}\phi_m\phi_nd\varOmega\frac{\partial \bm{Q}_m}{\partial t} &= \int_{\mathcal{D}^k}\left(\bm{F}(\bm{Q}^h)\frac{\partial \phi_n}{\partial x}+\bm{G}(\bm{Q}^h)\frac{\partial \phi_n}{\partial y}+\bm{H}(\bm{Q}^h)\frac{\partial \phi_n}{\partial z}\right)d\varOmega \\
&-\int_{\partial\mathcal{D}^k}\left[{n}_x\bm{F}(\bm{Q}^h)+{n}_y\bm{G}(\bm{Q}^h)+{n}_z\bm{H}(\bm{Q}^h)\right]^*\phi_nd\varGamma
\end{aligned}
\end{equation}

\noindent The integration term on the left-hand side is the component of the mass matrix, we have:
\begin{equation}
M_{mn}=\int_{\mathcal{D}_k}\phi_m(\bm{x})\phi_n(\bm{x})d\varOmega=\int_{\mathcal{I}'}\phi_m(\bm{\xi})\phi_n(\bm{\xi})\mathbb{J}(\bm{\xi})d\bm{\xi}
\end{equation}

\noindent where $\mathbb{J}(\bm{\xi})$ denotes the Jacobian determinant of the parametric transformation and can be expressed as follows:
\begin{equation}
\mathbb{J}(\bm{\xi})=\left|\frac{\partial(x,y,z)}{\partial(\xi,\eta,\zeta)}\right| =
\left|
\begin{array}{ccc}
\displaystyle\frac{\partial x}{\partial \xi} & \displaystyle\frac{\partial x}{\partial \eta} & \displaystyle\frac{\partial x}{\partial \zeta}\\
\displaystyle\frac{\partial y}{\partial \xi} & \displaystyle\frac{\partial y}{\partial \eta} & \displaystyle\frac{\partial y}{\partial \zeta}\\
\displaystyle\frac{\partial z}{\partial \xi} & \displaystyle\frac{\partial z}{\partial \eta} & \displaystyle\frac{\partial z}{\partial \zeta}
\end{array}
\right|
\end{equation}

\noindent Since the shape function for the parametric transformation of the triangle/tetrahedron is linear, the Jacobian determinant is constant in those elements. Therefore, we have $M_{mn}=\mathbb{J}\delta_{mn}$, which means that the mass matrix is diagonal. On the contrary, the shape function for the quadrilateral/hexahedron elements is bilinear, the Jacobian determinant is not a constant. To simplify the inversion of the mass matrix, for the quadrilateral/hexahedron elements, the mass matrix is rewritten by the row-sum lumped technique:
\begin{equation}
M_{mn} = \delta_{mn}\sum_{n}\int_{\mathcal{D}_k}\phi_m(\bm{x})\phi_n(\bm{x})d\varOmega
\end{equation}

\noindent Finally, the discretized governing equation can be rewritten as follows:
\begin{equation}
\label{eq-discretized-governing-equation}
\frac{\partial \bm{Q}_m}{\partial t}=M^{-1}_{mn}\cdot\left[\int_{\mathcal{D}^k}\left(\bm{F}\frac{\partial \phi_n}{\partial x}+\bm{G}\frac{\partial \phi_n}{\partial y}+\bm{H}\frac{\partial \phi_n}{\partial z}\right)d\varOmega - \int_{\partial\mathcal{D}^k}\left[{n}_x\bm{F}+{n}_y\bm{G}+{n}_z\bm{H}\right]^*\phi_nd\varGamma\right]
\end{equation}

\noindent in which, the right-hand side sums over the dummy index $n$. Our goal is to solve the coefficient $\bm{Q}_m$ by the above governing equation.

\subsection{Numerical flux}
In DGM, the state variables on the element boundary are multi-defined, which means that the fluid field is not asked to be continuous between two contact elements. To solve the numerical flux on the element boundary, the approximate Riemann solver devised by Harten, Lax, and van Leer (denoted HLL) is an appropriate choice for the current system governed by Euler equations. However, the single state HLL solver is too diffusive and unable to resolve isolated contact discontinuities well \cite{li2005hllc}. Therefore, the HLLC (HLL for contact wave) solver proposed by Toro \emph{et al.} \cite{toro1994restoration} is employed in the present work. The HLLC solver is expressed as follows:
\begin{equation}
\renewcommand{\arraystretch}{1.2}
\bm{F}^{HLLC}(\bm{Q}_L, \bm{Q}_R) = \left\{
\begin{array}{ll}
\bm{F}(\bm{Q}_L) & 0\le S_L \\
\bm{F}(\bm{Q}_L) + S_L(\bm{Q}_L^*-\bm{Q}_L) & S_L\le0\le S^* \\
\bm{F}(\bm{Q}_R) + S_R(\bm{Q}_R^*-\bm{Q}_R) & S^*\le0\le S_R \\
\bm{F}(\bm{Q}_R) & S_R\le 0 \\
\end{array}
\right.
\end{equation}

\noindent where $\bm{F}$ denotes the flux function, $\bm{Q}_L$ and $\bm{Q}_R$ are the initial state vector on the left and right sides of the element boundary, respectively. $\bm{Q}_L^*$ and $\bm{Q}^*_R$ are the intermediate states, while the intermediate region is separated into left and right parts by the contact wave. $S_L$ and $S_R$ denote the slowest and fastest wave speed, respectively, $S^*$ indicates the middle wave speed, and we have:
\begin{equation}
\begin{aligned}
S_L &= {\rm min}(u_L-c_L, u_R-c_R) , \,\, S_R = {\rm max}(u_L+c_L, u_R+c_R) \\
S^* &= \frac{p_R-p_L+\rho_Lu_L(S_L-u_L)-\rho_Ru_R(S_R-u_R)}{\rho_L(S_L-u_L)-\rho_R(S_R-u_R)}
\end{aligned}
\end{equation}

\noindent where $u$, $p$, and $\rho$ denote the fluid velocity, pressure, and density, respectively. $c_L$ and $c_R$ are local sound speeds on the left and right sides of the element boundary, respectively. For the multi-dimension problem, the velocity is transformed into the local coordinate system. The basis vectors of the local coordinate system are the boundary outward normal and two tangent vectors. And the transform matrix $\bm{q}$ between the global and local coordinate system is related to the basis vectors. We have:
\begin{equation}
q_{ij} = \bm{E}^l_i\cdot \bm{E}^g_j \text{,  in which: } \bm{E}_1^l=\bm{n}\text{, }\,\bm{E}_2^l=\bm{t}_1\text{, }\,\bm{E}_3^l=\bm{t}_2
\end{equation}

\noindent where $\bm{E}^l$ and $\bm{E}^g$ are the basis vectors of the local and global coordinate system, $\bm{n}$ is the boundary unit normal, $\bm{t}_1$ and $\bm{t}_2$ are two unit tangent vectors perpendicular to each other. Thus, the velocity in the local coordinate system $\bm{v}^l=\bm{q}\cdot\bm{v}^g$, where $\bm{v}^g$ denotes the velocity in the global coordinate system. The velocity components in all the above variables should be replaced by $\bm{v}^l = (u^l,v^l,w^l)$, e.g., $\bm{Q}_L=[\rho_L, \rho_Lu_L^l, \rho_Lv_L^l, \rho_Lw_L^l, E_L]^{\rm T}$. And the intermediate state is expressed as follows: 
\begin{equation}
\bm{Q}^*_I = \rho_I\frac{S_I-u^l_I}{S_I-S^*}\left[
\renewcommand{\arraystretch}{1.0}
\begin{array}{c}
1 \\
S^* \\
v^l_I \\
w^l_I \\
\displaystyle \frac{E_I}{\rho_I}+(S^*-u^l_I)\left(S^* + \frac{p_I}{\rho_I(S_I-u^l_I)} \right)
\end{array}
\right]
\end{equation}

\noindent After calculating the HLLC flux in the local coordinate system, we need to transform the flux back into the global coordinate system. The algorithm is expressed as follows:
\begin{equation}
\begin{aligned}
\hat{F}^*_1 &= F^{HLLC}_1\text{, } \hat{F}^*_5 = F^{HLLC}_5\\
\hat{F}^*_{i+1} &= q^{-1}_{ij}F^{HLLC}_{j+1} \text{, in which: } i,j\in\left\{1,2,3\right\}
\end{aligned}
\end{equation}

\noindent where $q_{ij}^{-1}$ is the component of the inverse of the transform matrix $\bm{q}$, let the numerical flux $[n_x\bm{F}+n_y\bm{G}+n_z\bm{H}]^* = \hat{\bm{F}}^*$, and substitute it into the governing equation Eq. \eqref{eq-discretized-governing-equation}. Afterward, the solution of the governing equation is ready.

\section{The solution of the governing equation}
\subsection{Integration over the element}
To solve Eq. \eqref{eq-discretized-governing-equation}, the terms at the right-hand side should be calculated first. In the present work, the Gauss-Jacobi quadrature is employed to obtain the numerical integration in Eq. \eqref{eq-discretized-governing-equation}. In one dimension, The Gauss-Jacobi quadrature rule is defined as follows:
\begin{equation}
\int_{-1}^{1}\omega^{(\alpha,\beta)}(\xi)\cdot f(\xi)d\xi\approx\sum_{i=1}^{n}w_{i}f\left(\xi_{i}\right)
\end{equation}

\noindent where $f(\xi)$ is a smooth function on $[-1,1]$, and $\omega^{(\alpha,\beta)}(\xi)=(1-\xi)^{\alpha}(1+\xi)^{\beta}$ is the weight function. $w_i$ is the integration weight of the $i$-th integration point, $f(\xi_i)$ is the value of the function at $\xi_i$, and $\xi_i$ is the coordinate of the $i$-th integration point. The Gauss-Jacobi quadrature is related to the Jacobi polynomials. For instance, the integration points $\{\xi_1,\xi_2,...,\xi_n\}$ are the roots of the $n$-th order Jacobi polynomial $J^{(\alpha,\beta)}_n(\xi)$. For the calculation of integration points, Golub and Welsch's research \cite{golub1969calculation} shows that $\xi_i$ are the eigenvalues of the recurrence equations that truncated by $J^{(\alpha,\beta)}_n(\xi)=0$. The integration weight of the $i$-th integration point is given by the formula \cite{hale2013fast}:
\begin{equation}
w_{i}=\frac{2^{\alpha+\beta+1}\Gamma(\alpha+n+1)\Gamma(\beta+n+1)}{n!\Gamma(\alpha+\beta+n+1)\left(1-\xi_{i}^{2}\right)\left[J_{n}^{(\alpha,\beta)^{\prime}}(\xi_{i})\right]^{2}}
\end{equation}

\noindent The roots of the Jacobi polynomial and the normalized Jacobi polynomial are the same. Considering the derivative of the normalized Jacobi polynomial Eq. \eqref{eq-derivative-jacobi-poly}, and combine with the Eq. \eqref{eq-basis-function}, the above equation can be expressed as:
\begin{equation}
w_{i}=\frac{2n+\alpha+\beta+1}{n(n+\alpha+\beta+1)\left(1-\xi_{i}^{2}\right)\left[P_{n-1}^{(\alpha+1,\beta+1)}(\xi_{i})\right]^{2}}
\end{equation}

\noindent Substituting $\alpha=\beta=0$ into the above equation, which can be simplified as follows:
\begin{equation}
w_{i}=\frac{2n+1}{n(n+1)\left(1-\xi_{i}^{2}\right)\left[P_{n-1}^{(1,1)}(\xi_{i})\right]^{2}}
\end{equation}

\noindent The Gauss-Jacobi quadrature rule with $n$ integration points can yield an exact integration result for polynomials $f(\xi)$ of degree $2n-1$ or less. While the coordinates and weights for the one-dimensional Gauss-Jacobi quadrature are obtained, for the 2D square or 3D cube isoparametric elements, the numerical integration can be obtained by the tensor product. For instance, we consider the integration inside the quadrilateral element, which can be transformed onto the standard square (isoparametric element), the numerical integration over the 2D standard square can be calculated as follows:
\begin{equation}
\int_{\mathcal{D}^k}f(x,y)dxdy=\int_{-1}^{1}\int_{-1}^{1}f(\xi,\eta)\mathbb{J}(\xi,\eta)d\xi d\eta=\sum_{i=1,j=1}^{n\times n}w_iw_jf(\xi_i,\eta_j)\mathbb{J}(\xi_i,\eta_j)
\end{equation}

\noindent where $\mathbb{J}(\xi_i,\eta_j)$ denotes the Jacobian determinant at the integration point $(\xi_i,\eta_j)$. $\xi_i$ and $\eta_j$ denote the coordinates in the horizontal and vertical direction, respectively, and both are obtained by the one-dimensional Gauss-Jacobi quadrature. $w_i$ and $w_j$ are the corresponding integration weights. In the present work, for the numerical integration on the triangle and tetrahedron isoparametric elements, the coordinates and weights are calculated by the Grundmann-M{\"o}ller rule \cite{grundmann1978invariant}. Using the Gauss-Jacobi quadrature, the coefficient in the modal expansion Eq. \eqref{eq-modal-expandsion-coefficient} can be obtained:
\begin{equation}
\bm{Q}_n=\sum_{i=1}^{N_g}w_i\bm{Q}^h(\bm{\xi}_i,t)\phi_n(\bm{\xi}_i)\mathbb{J}(\bm{\xi}_i)
\end{equation}
 
\noindent where $N_g$ is the total number of the integration point in the current element. The above equation will be used to initialize the state variables at the beginning of the simulation.

\subsection{Integration over the element boundary}
\begin{figure}[htpb]
\centering
\includegraphics[width=4.4in]{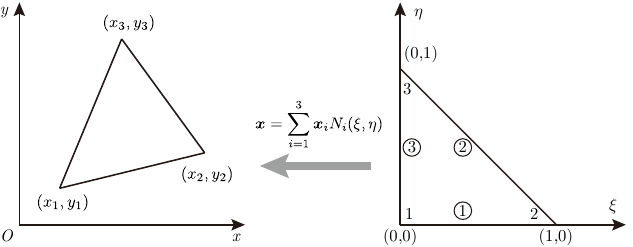}
\caption{The mapping for triangle element. $\bm{x}_i$ denotes the coordinates of the element vertexs, $N_i(\xi,\eta)$ is the FEM shape function for triangular element.}
\label{fig-triangle-mapping}
\end{figure}
For the integration on the element boundary in $n$ dimension, the coordinates and weights of the boundary integration points are obtained by the $n-1$ dimensional Gauss quadrature. For instance, the boundary integration of triangle elements can be obtained by adopting the one-dimensional Gauss quadrature. Specifically, we map the original triangle to the standard triangle $\mathcal{I}=\{(\xi,\eta)|\xi,\eta\ge0, \xi+\eta\le1\}$, as shown in Fig. \ref{fig-triangle-mapping}, therefore, the coordinates of the boundary integration points are:
\begin{equation}
\renewcommand{\arraystretch}{1.0}
\left\{
\begin{array}{lll}
\text{On edge 1:} & \xi_i=0.5(1+\xi^{1D}_i) & \eta_i=0 \\
\text{On edge 2:} & \xi_i=0.5(1+\xi^{1D}_i) & \eta_i=1-\xi_i \\
\text{On edge 3:} & \xi_i=0 &  \displaystyle \eta_i=0.5(1+\xi^{1D}_i)
\end{array}
\right.
\end{equation}

\noindent where $\xi^{1D}_i$ are the coordinates of the one-dimension Gauss quadrature points and the edge index is shown in Fig. \ref{fig-triangle-mapping}. Besides, the weights on each edge are the same as the weights in one dimension, which means that $w_i=w_i^{1D}$. In the governing equation Eq. \eqref{eq-discretized-governing-equation}, to calculate the boundary integration term on the right-hand side, firstly, we need to obtain the Jacobian determinant for the mapping along the boundary; Secondly, we need to calculate the boundary normal. For more details, readers may refer to one of the most excellent textbooks of the Discontinuous Galerkin Method \cite{hesthaven2007nodal}.

\subsection{Time integration}
In the above sections, the right-hand side of the governing equation Eq. \eqref{eq-discretized-governing-equation} can be determined. To simplify the description, the governing equation is rewritten as follows:
\begin{equation}
\frac{\partial \bm{Q}_m}{\partial t} = \mathcal{R}_m(\bm{Q},t)
\end{equation}

\noindent where $\bm{Q}_m$ denotes the $m$-th coefficient of the modal expansion Eq. \eqref{eq-modal-expandsion}, and $\mathcal{R}_m(\bm{Q}^h,t)$ denotes the value of right hand side of the governing equation Eq. \eqref{eq-discretized-governing-equation}. The above equation is a system of ordinary differential equations. In the present work, the optimal Runge-Kutta integration scheme has been adopted. The order of Runge-Kutta should be higher than the order of the orthogonal polynomials. For instance, when the highest polynomial order is 2, the optimal third-order three-stage strong stability-preserving Runge-Kutta (SSP-RK) scheme is adopted to solve the state variable $\bm{Q}$. The algorithm is expressed as follows:
\begin{equation}
\begin{aligned}
\bm{Q}^{(1)}_m &= \bm{Q}^{t}_m + \Delta t\mathcal{R}_m(\bm{Q}^t,t) \\
\bm{Q}^{(2)}_m &= \frac{1}{4}\left(3\bm{Q}^t_m+\bm{Q}^{(1)}_m+\Delta t\mathcal{R}_m(\bm{Q}^{(1)},t+\Delta t)\right) \\
\bm{Q}^{t+\Delta t}_m &= \frac{1}{3}\left(\bm{Q}^t_m+2\bm{Q}^{(2)}_m+2\Delta t\mathcal{R}_m(\bm{Q}^{(2)},t+\frac{1}{2}\Delta t)\right) \\
\end{aligned}
\end{equation}

\noindent For the unsteady flow problem studied in this paper, the global time increment is applied to the time integration of the local element. To ensure numerical stability, the time increment needs to satisfy the following Courant-Friedrichs-levy (CFL) condition:
\begin{equation}
\Delta t= C \left(\frac{h}{\|\bm{u}\|+c}\right)_{\rm min}
\end{equation}

\noindent where $h$ denotes the length of the shortest edge of the element, $\bm{u}$ is the fluid velocity, $c$ denotes the speed of sound, and $C$ is the CFL number. In the present work, we set $C\le 0.4/(2N+1)$ to ensure the time integration is linearly stable \cite{luo2007hermite}, in which $N$ is the highest order of the polynomial in the modal expansion.

\subsection{Shock detection and limiter}
In the compressible fluid flow problem, shock waves may appear in the flow field and the fluid variables will be discontinuous near these areas. The high-order numerical scheme can obtain a more accurate result in the smooth area, however, it will bring spurious oscillations when dealing with strong discontinuities. To reduce or remove the spurious oscillations near the discontinuities, many limiters have been developed. In numerical practice, it will reduce the solution accuracy if the limiter is used in the smooth fluid domain. Therefore, we should detect the troubled cell in the discontinuous area that need the limiting procedure, and then replace the solution polynomial with the reconstructed polynomial. In the present work, the trouble cell indicator developed by Fu and Shu \cite{fu2017new} is adopted, which can capture shocks without PDE-sensitive parameters. While the troubled cell is detected, the vertex-based slope limiter proposed by Kuzmin \cite{kuzmin2010vertex} is applied to reconstruct the fluid conservative variables.

\section{Convergence rate verification}
In this section, the two dimensional Euler Equations is solvered with the initial condition defined as $\rho(x,y)=1.0+0.2\sin(\pi(x+y))$, $u(x,y)=0.7$, $v(x,y)=0.3$ and $p(x,y)=1.0$. Where $\rho$ denotes the fluid density, $u$ and $v$ are the velocity in $x$ and $y$ directions, respectively. $p$ represents the fluid pressure, and we have $p=(\gamma-1)(E-\frac{1}{2}\rho(u^2 + v^2))$, with specific heat ratio $\gamma$ taken as 1.4.
\begin{figure}[htpb]
\centering
\includegraphics[width=5.0in]{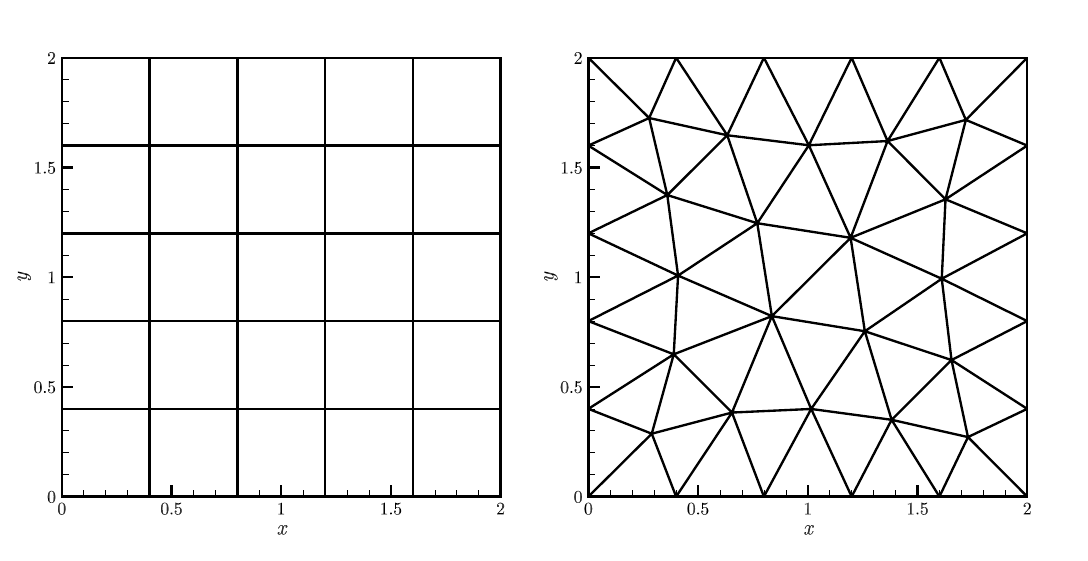}
\caption{Meshes for the simulation of 2D Euler equations, $h=4/10$. Left: quadrilateral meshes; Right: triangular meshes.}
\label{fig-accuracy-verification-mesh}
\end{figure}

The computational domain is $[0,2]\times[0,2]$, and periodic boundary conditions are applied on each side of the domain. The total simulation time is $t=2.0$. Both quadrilateral and triangular mesh are tested, and the coarsest mesh quadrilateral meshes and triangular meshes are shown in Fig. \ref{fig-accuracy-verification-mesh}. In the simulation, the cell length $h$ varies from $4/10$ to $4/160$, and the highest order of the orthogonal polynomial varies from 1 to 3. The $L^1$ and $L^2$ errors are calculated to verify the accuracy of the current solver, which is defined as follows:
\begin{equation}
L^1 = \frac{1}{N}\sum_{i=1}^{N} |\rho_i-\rho_{\rm exact}| ,\,\,\, L^2 = \sqrt{\frac{1}{N}\sum_{i=1}^{N}(\rho_i-\rho_{\rm exact})^2}
\end{equation}

\noindent here $\rho_i$ represents the numerical simulaition element quadrature point value, varys from element type and polynomial degree. $\rho_{\rm exact}$ is the analytical solution value, and we have $\rho_{\rm exact}(x,y)=1.0+0.2\sin(\pi(x+y-t))$. $N$ denotes the total number of the quadrature points inside the elements.

The computational error at the final step is shown in Table \ref{tab:DG-accuracy-verification}. In which is the convergence rate is calculated, and the CPU time costs are given. From the result, we can seen that the error decreases with the reduction of the cell length and the increase of the polynomial degree. The convergence rate is close to the analytical value, which indicates the correctness of the numerical scheme. Meanwhile, it has been found that high-order schemes can achieve very small errors even when using coarser grids. Therefore, for the same level of error, the computational time of high-order schemes can be even shorter. Consequently, we should not believe that high-order algorithms inherently have low computational efficiency. Additionally, the error for triangular meshes is slightly larger than that for quadrilateral meshes, which is due to the fact that triangular meshes typically have a larger aspect ratio, leading to higher numerical error.
\begin{table}[h]
\tiny
\centering
\caption{2D-Euler equations:initial data $\rho(x,y)=1.0+0.2\sin(\pi(x+y))$, $u(x,y)=0.7$, $v(x,y)=0.3$ and $p(x,y)=1.0$}
\label{tab:DG-accuracy-verification}
\renewcommand{\arraystretch}{1.0} 
\resizebox{\textwidth}{!}{ 
\scalebox{0.7}{
\begin{tabular}{cc ccccc ccccc}
\toprule
$p$     & $h$   & \multicolumn{5}{c}{Quadrilateral elements}   & \multicolumn{5}{c}{Triangular elements}    \\
\cmidrule(r){3-7} \cmidrule(r){8-12}
        &       &$L^1$ error & Order & $L^2$ error  & Order &CPU Time(s)  &$L^1$ error & Order & $L^2$ error & Order &CPU Time(s)    \\
\toprule
$p^1$   & 4/10  & 1.39E-02	&    	&1.73E-02	&      &0.02191	&2.25E-02	&    	&2.64E-02	&      &0.03506	\\
        & 4/20  & 1.89E-03	&2.88 	&2.45E-03	&2.82  &0.05262	&2.86E-03	&2.98 	&3.87E-03	&2.77  &0.09567	\\
        & 4/40  & 3.88E-04	&2.28 	&4.94E-04	&2.31  &0.2290 	&4.78E-04	&2.58 	&6.54E-04	&2.56  &0.6348	\\
        & 4/80  & 1.08E-04	&1.85 	&1.27E-04	&1.96  &1.543 	&1.09E-04	&2.13 	&1.44E-04	&2.18  &3.491	\\
        & 4/160 & 2.40E-05	&2.17 	&2.86E-05	&2.15  &11.72 	&2.66E-05	&2.04 	&3.42E-05	&2.07  &21.93	\\
\midrule          			
$p^2$   & 4/10  & 1.54E-03	&    	&1.92E-03	&      &0.04387	&1.95E-03	&    	&2.36E-03	&      &0.08318	\\
        & 4/20  & 2.88E-04	&2.42 	&3.52E-04	&2.45  &0.1369 	&2.71E-04	&2.85 	&3.35E-04	&2.82  &0.2211	\\
        & 4/40  & 4.85E-05	&2.57 	&5.84E-05	&2.59  &0.7305 	&3.16E-05	&3.10 	&3.93E-05	&3.09  &1.703	\\
        & 4/80  & 6.90E-06	&2.81 	&8.28E-06	&2.82  &5.559 	&3.78E-06	&3.06 	&4.73E-06	&3.06  &11.44	\\
        & 4/160 & 8.99E-07	&2.94 	&1.08E-06	&2.94  &33.73 	&4.53E-07	&3.06 	&5.70E-07	&3.05  &77.74	\\
\midrule          			
$p^3$   & 4/10  & 7.46E-05	&    	&9.07E-05	&      &0.09001	&1.77E-04	&    	&2.31E-04	&      &0.1742	\\
        & 4/20  & 2.94E-06	&4.67 	&3.52E-06	&4.69  &0.4848	&8.89E-06	&4.32 	&1.21E-05	&4.26  &1.285	\\
        & 4/40  & 1.64E-07	&4.17 	&1.99E-07	&4.14  &3.080 	&4.55E-07	&4.29 	&6.08E-07	&4.31  &7.217	\\
        & 4/80  & 9.87E-09	&4.05 	&1.21E-08	&4.04  &21.21 	&2.64E-08	&4.11 	&3.54E-08	&4.10  &41.93	\\
        & 4/160 & 6.77E-10	&3.86 	&8.70E-10	&3.80  &150.9 	&1.56E-09	&4.08 	&2.09E-09	&4.08  &306.6	\\
\bottomrule
\end{tabular}}}
\end{table}

\section{Numerical examples}
\subsection{Forward step problem}
\begin{figure}[htpb]
\centering
\includegraphics[width=6.6in]{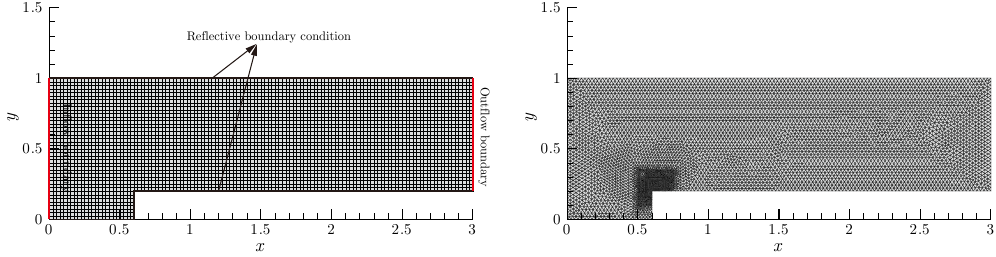}
\caption{Forward step problem. Left: uniform distributed quadrilateral mesh with cell length $h=1/40$; Right: unstructured triangle mesh with cell length $h$ vary in $\left[1/40, 1/160 \right]$.}
\label{fig-forward-step-problem-set}
\end{figure}
In this section, the Mach 3 flow in the wind tunnel with a step is simulated. This problem comes from \cite{woodward1984numerical}. The dimension of the problem is shown in Fig. \ref{fig-forward-step-problem-set}, the length and width of the wind tunnel are 3 and 1, respectively. The step height is 0.2 and 0.6 from the left boundary of the wind tunnel. In the simulation, the upper and lower wall of the tunnel is set as reflective boundary conditions. And the left and right boundaries are set as inflow and outflow boundary conditions. At the beginning of the simulation, airflow into the wind tunnel at Mach 3. The simulation lasts to $t=4$.
\begin{figure}[htpb]
\centering
\includegraphics[width=6.0in]{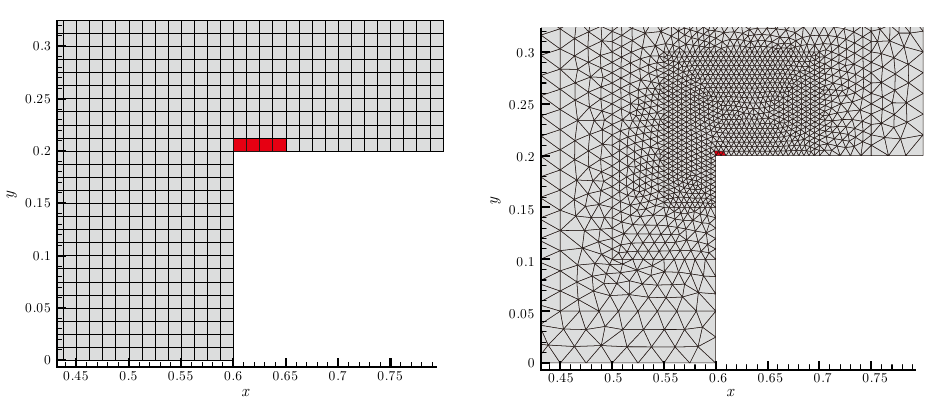}
\caption{The fluid variables are reset on the elements, which are marked red. Left: the uniform quadrilateral mesh while cell length $h=1/80$; Right: unstructured triangle mesh with cell length $h$ vary in $[1/40,1/160]$.}
\label{fig-forward-step-problem-singularity}
\end{figure}

The corner of the step is a singularity. To avoid the formation of a spurious entropy layer, the additional boundary condition near the step corner proposed by Woodward and Colella \cite{woodward1984numerical} is adopted. The fluid variable in the above corner is reset. These elements are marked red and shown in Fig. \ref{fig-forward-step-problem-singularity}. In these elements, we reset density to ensure the entropy value ($S=c_v\ln\frac{p}{\rho^\gamma}$) remains the same as the value in the lower left corner of the step. Besides, the velocity magnitudes are also reset to ensure the kinetic energy per unit mass in these elements has the same value as the lower left corner of the step.
\begin{figure}[htpb]
\centering
\includegraphics[width=6.6in]{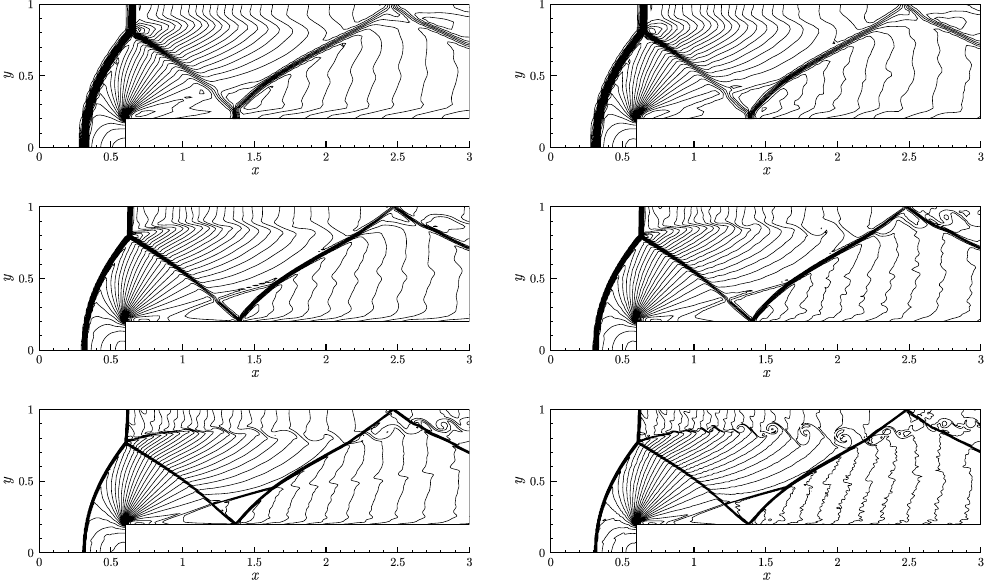}
\caption{Density contours of the fluid domain. The fluid density changes from $0.32$ to $6.0$ on 30 contours. Top to bottom: the cell length of the uniform quadrilateral meshes are $1/40$, $1/80$, and $1/160$. Left: second order, the highest polynomial order $N=1$; Right: third order $N=2$.}
\label{fig-forward-step-rho-contour-quad}
\end{figure}

\begin{figure}[htpb]
\centering
\includegraphics[width=6.6in]{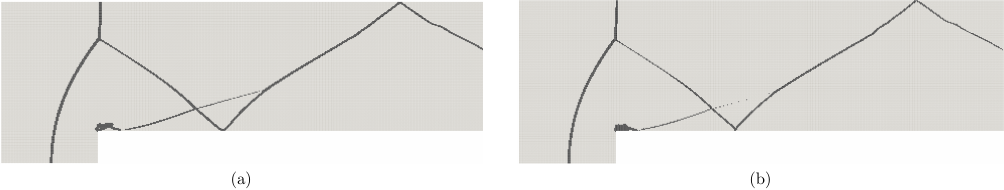}
\caption{Forward step problem. Troubled cells are marked black, the highest polynomial order (a) $N=1$, in (b) $N=2$. The mesh points on the boundary are uniformly distributed with cell length h = 1/80. The threshold for the troubled-cell indicator (2.4) is $C_k = 0.015 \times 2k -1 $.}
\label{fig-forward-step-trouble-cell}
\end{figure}

The density contours calculated by the quadrilateral mesh are shown in Fig. \ref{fig-forward-step-rho-contour-quad}. It can be seen that the calculated shock wave interface is sharper, and more details of the fluid flow can be captured with the reduction of the cell length. Besides, when the cell length is the same, comparing the results of $N=1$ and $N=2$, it can be seen that increasing the order of the polynomial can capture more flow structures. Besides, the discontinuous captured by the trouble-cell indicator is presented in Fig. \ref{fig-forward-step-trouble-cell}, where the troubled cell is marked black. From this, it can be seen that the captured discontinuous areas match well with the density contour results, and no elements in the smoothed fluid domain are marked, indicating the correctness of the discontinuous capture scheme implementation.

\begin{figure}[htpb]
\centering
\includegraphics[width=6.6in]{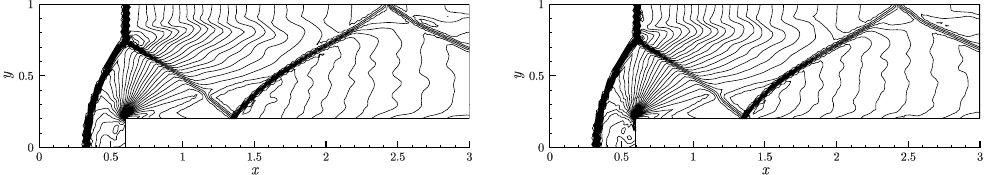}
\caption{Density contours of the fluid domain that discretized by the unstructured triangle element. The fluid density changes from $0.32$ to $6.0$ on 30 contours. The cell length of the triangle mesh varies in $[1/40, 1/160]$. Left: second order $N=1$; Right: third order $N=2$.}
\label{fig-forward-step-rho-contour-tri}
\end{figure}

Finally, the density contours obtained using triangular elements are shown in Fig. \ref{fig-forward-step-rho-contour-tri}. This figure demonstrates that the distribution of shock wave interfaces aligns well with the results obtained using quadrilateral elements. Additionally, due to the low mesh resolution at the shock wave interface, there is little variation in the calculation results across different order solvers, as seen in the density contour results.

\subsection{Double Mach reflection}
In this section, we simulate the double Mach reflection problem. This problem is originally from \cite{woodward1984numerical}. The computational domain for this problem is $[0,4]\times[0,1]$. The reflecting wall lies at the bottom, starting from $\displaystyle x=\frac{1}{6}$. Initially, a right-moving Mach 10 shock is positioned at $\displaystyle x=\frac{1}{6}$, $y=0$ and makes a 60$^\circ$ angle with the x-axis. For the bottom boundary, the exact post-shock condition is imposed for the part from $x=0$ to $\displaystyle x=\frac{1}{6}$ and a reflective boundary condition is used for the rest. At the top boundary, the flow values are set to describe the exact motion of a Mach 10 shock. In the simulation, quadrilateral elements are adopted, and the mesh sizes are chosen as $1/60$, $1/120$ and $1/240$. The total simulation time is set as $t=0.2$.
\begin{figure}[htpb]
\centering
\includegraphics[width=6.6in]{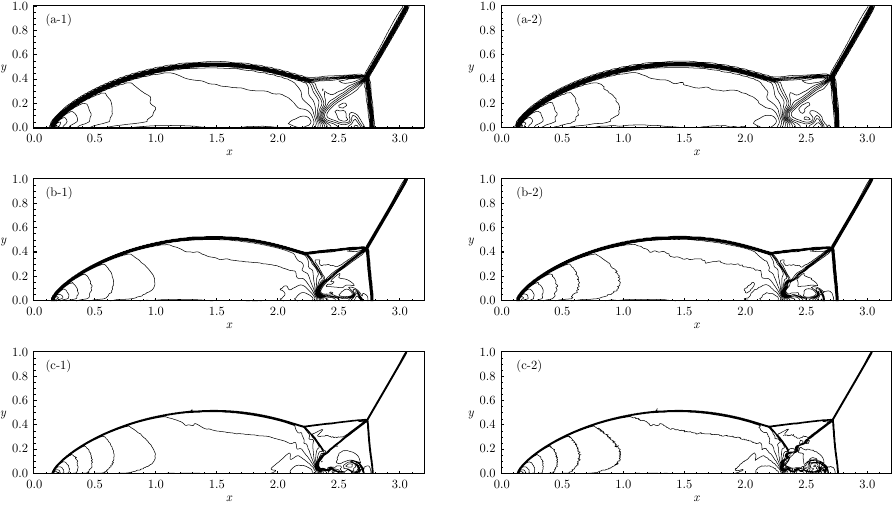}
\caption{Double Mach reflection problem. The fluid density changes from 0.562 to 20.9 on 30 contours. Top to bottom: the mesh size are $1/60$, $1/120$ and $1/240$. Left: second order, the highest polynomial order, $N=1$; Right: third order, $N=2$.}
\label{figure_DifferentOrderAndDifferentMeshes}
\end{figure}

\begin{figure}[htpb]
\centering
\includegraphics[width=6.6in]{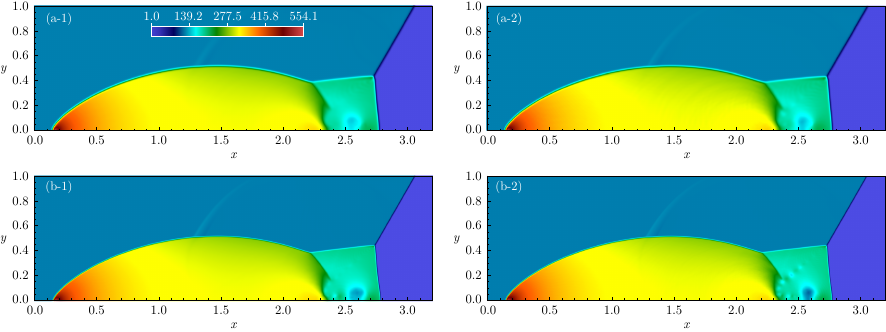}
\caption{Fluid pressure obtained by different mesh size (top: $\Delta x = 1/120$; bottom: $\Delta x = 1/240$). Left: polynomial order $N=1$; Right: polynomial order $N=2$.}
\label{WB-Euler_mach_reflect4n_240H_2order}
\end{figure}

This problem is simulated using different polynomial orders and mesh sizes. The fluid density contour results are shown in Fig. \ref{figure_DifferentOrderAndDifferentMeshes}, from which it can be seen that the shock interface is consistent with the results in the literature \cite{woodward1984numerical}. Additionally, increasing the mesh resolution and polynomial order allows for more detailed simulation results, enabling better capture of shock wave structures, particularly near the intersections of multiple shock waves. Finally, the fluid pressure is shown in Fig. \ref{WB-Euler_mach_reflect4n_240H_2order}. This indicates that the calculated wavefront is very sharp, with clear discontinuities in the pressure field before and after the wavefront, while the pressure distribution in the smooth areas remains quite smooth.

\subsection{Transonic flow past a NACA0012 airfoil}
We consider a two-dimensional inviscid Euler supersonic flow with a free-stream Mach number of $M_\infty=0.8$ over a NACA0012 airfoil at an angle of attack of 1.25 degrees. The chord length of the airfoil is 1. The computational domain spans $[-4,-4]\times[6,4]$, with the leading edge of the airfoil positioned at (0,0). The computational grid is a non-uniform grid, with a cell size of 0.004 around the airfoil and 0.2 at the outer boundary, consisting of a total of 29320 quadrilateral cells. The left boundary has an inflow Mach number of 0.85, and the right boundary has an outflow. To eliminate reflection, the top and bottom boundaries are also set as the inflow boundary. The airfoil surface is treated as a free-slip wall boundary. The total simulation time is set as $t=40$.
\begin{figure}[htpb]
\centering
\includegraphics[width=6.6in]{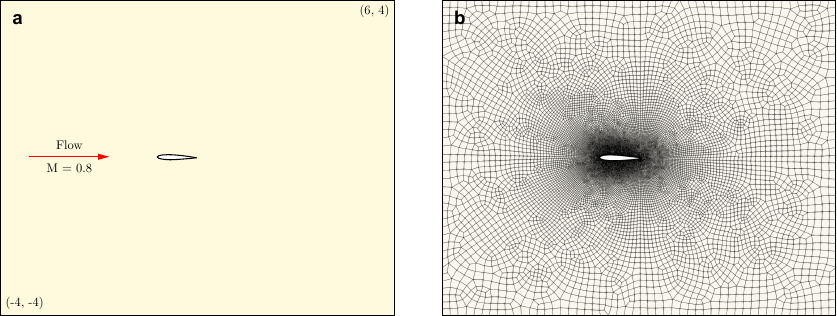}
\caption{NACA0012 arifoil Schematic Diagram.}
\label{SchardinProblemSchematicDiagram}
\end{figure}

\begin{figure}[htpb]
\centering
\includegraphics[width=6.6in]{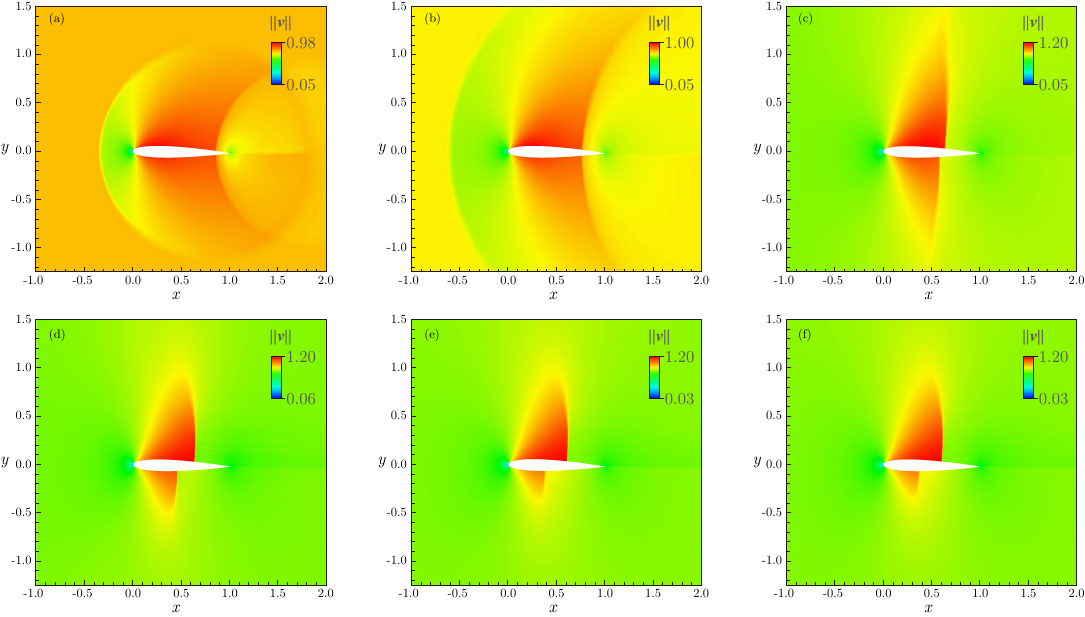}
\caption{Velocity contour history calculated using the present DG solver with $N=1$: (a) $t=1$; (b) $t=2$; (c) $t=5$; (d) $t=10$; (e) $t=30$; (f) $t=40$.}
\label{SchardinProblemVelContour}
\end{figure}

\begin{figure}[htbp]
\centering
\begin{minipage}{0.49\textwidth}
\centering
\includegraphics[width=3.0in]{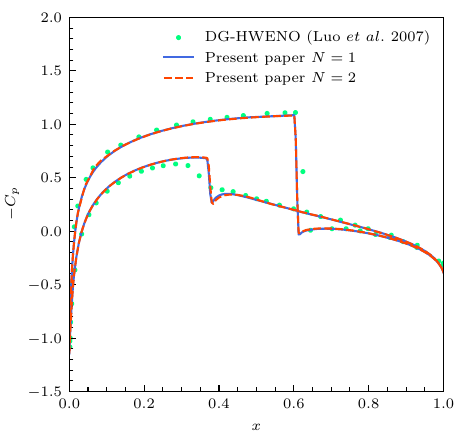}
\caption{Pressure Coefficient Distribution along the Airfoil, the result in \cite{luo2007hermite} is plotted as reference.}
\label{NACA0012}
\end{minipage}
\hfill
\begin{minipage}{0.49\textwidth}
\centering
\includegraphics[width=3.0in]{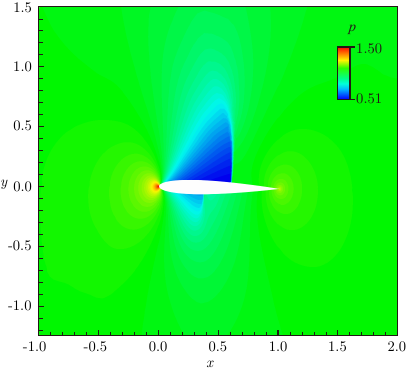}
\caption{Pressure contours at the end of the simulation, $N=1$.}
\label{NACA0012_pressure}
\end{minipage}
\end{figure}

The time evolution of the velocity contour lines is shown in the Fig. \ref{SchardinProblemVelContour}. Initially, two shock waves form at the leading and trailing edges of the airfoil. As the simulation progresses, the shock waves gradually stabilize in their final positions. The similarity between the results at $t=30$ and $t=40$ demonstrates the stability of the solution presented in this study. The pressure coefficient distribution is shown in Figure \ref{NACA0012}, with the results from \cite{luo2007hermite} plotted as a reference. From this, we can see that the pressures obtained by the first-order and second-order methods are basically consistent. Additionally, the mesh size in the present work is smaller than that in the literature, resulting in sharper shock interfaces. The pressure contours is shown in Figure \ref{NACA0012_pressure}, where a strong shock wave can be observed on the upper surface and a weak shock wave on the lower surface.

\subsection{Schardin's problem}
Schardin's problem was originally introduced in \cite{schardin1957high}. In 2000, \cite{chang2000shock} revisited the experiment, improving the setup to reduce noise. In this section, we simulate Schardin's experiment based on the work in \cite{chang2000shock}. The computational domain is defined as $x\in(-0.05, 0.092)$, $y\in(-0.075, 0.075)$. The lateral length of the equilateral triangular obstacle is 20 mm, with its vertices located at $(0,0)$, $(0.017, 0.01)$, $(0.017, -0.01)$. Both the triangular obstacle boundary and the domain boundary are modeled as free-slip reflective boundaries. The left boundary serves as an inflow boundary, while the right boundary acts as an outflow boundary. The upper and lower boundaries are also free-slip reflective. The shock wave propagates from left to right.
\begin{figure}[htpb]
\centering
\includegraphics[width=6.6in]{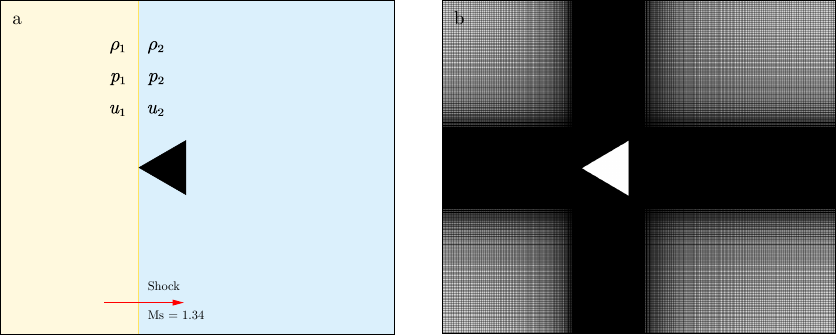}
\caption{Schardin's problem Schematic Diagram.}
\label{SchardinProblemSchematicDiagram}
\end{figure}

The static pressure on the low-pressure side of the shock tube is $\rho_2 = 1.2225 {\rm kg/m}^3$, $p_2=101325{\rm Pa}$. The initial shock wave is positioned at $x=0$, with a Mach number $M_s=1.34$. Behind the wave, the fluid density $\rho_1=1.938 {\rm kg/m}^3$ the pressure is set to $p_1=195374.86 {\rm Pa}$, and the velocity is $u_1=168.54 {\rm m/s}$. Ahead of the wave, fluid is at static state, the velocity is 0 m/s. The specific heat ratio is $\gamma=1.4$. The total simulation time is set to 0.2 ms. Mesh size near the triangular obstacle is 0.05 mm, and at the edges of the computational domain, it is 0.1 mm. The entire computational domain uses a gradient mesh to ensure resolution adapts to the geometry. The problem discription and mesh are shown in Fig. \ref{SchardinProblemSchematicDiagram}, total number of the quadridual elements are 581606.

\begin{figure}[htpb]
\centering
\includegraphics[width=6.6in]{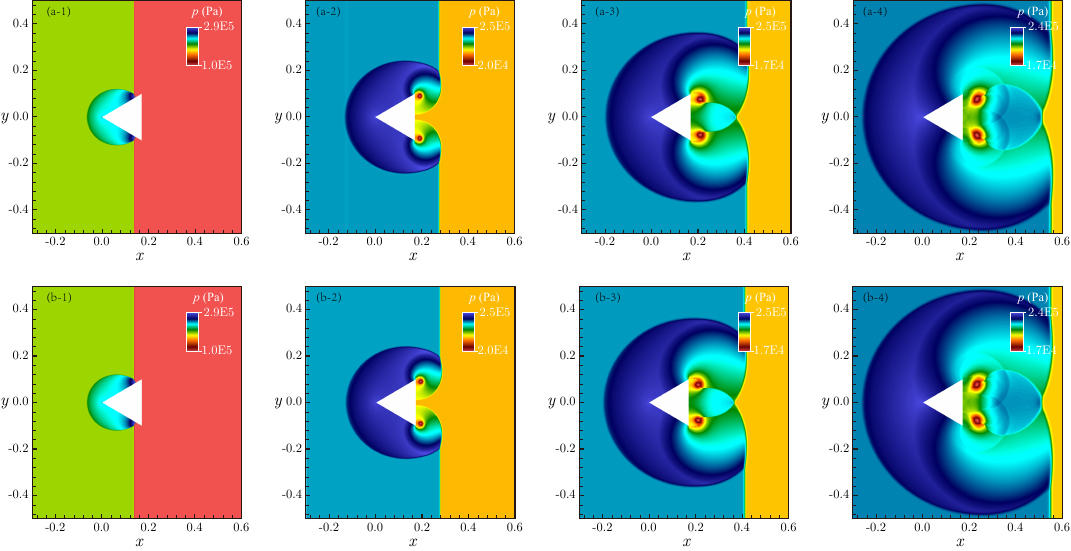}
\caption{The pressure contour history obtained by the present DGM solver, upper: $N=1$, lower: $N=2$. From left to right, the times are $t=30\mu{\rm s}$, $60\mu{\rm s}$, $90\mu{\rm s}$, and $120\mu{\rm s}$, respectively.}
\label{SchardinProblemPressure}
\end{figure}

\begin{figure}[htpb]
\centering
\includegraphics[width=6.6in]{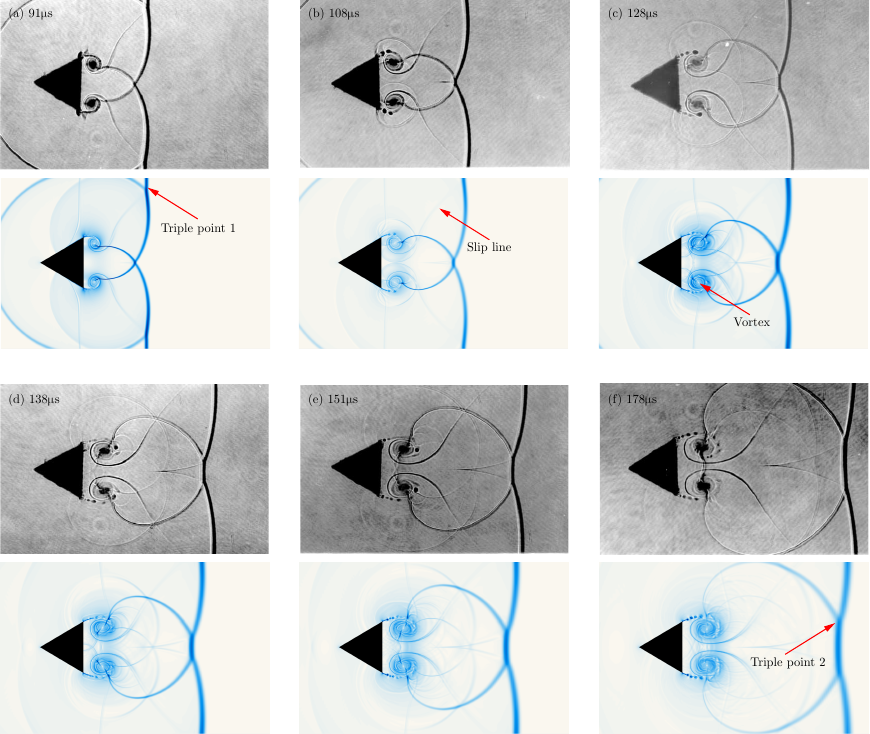}
\caption{Schardin's Problem: Comparison of Experimental and Numerical Results}
\label{Schileren_Comparison}
\end{figure}

\begin{figure}[htbp]
\centering
\begin{minipage}{0.45\textwidth}
\centering
\includegraphics[width=3.0in]{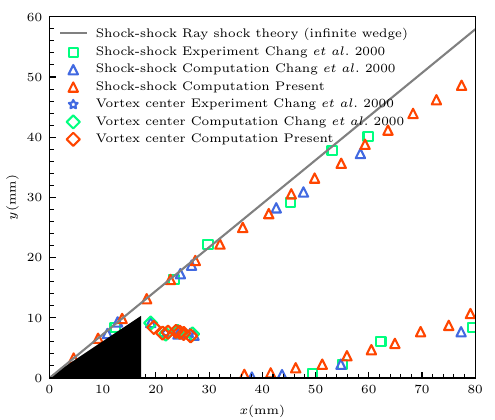}
\caption{The triple point trajectory of the shock wave and vortex center trajectory.}
\label{SchardinProblem_trajectory}
\end{minipage}
\hfill
\begin{minipage}{0.45\textwidth}
\centering
\includegraphics[width=3.0in]{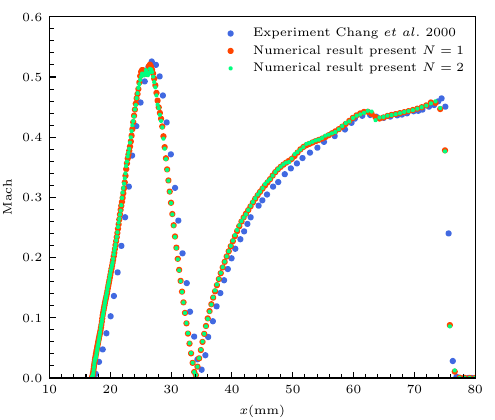}
\caption{The Mach distribution along the x-axis at time $t=172 \mathrm{\mu s}$.}
\label{SchardinProblem_Mach_distribution}
\end{minipage}
\end{figure}

The evolution of the prssure field is presented in Fig. \ref{SchardinProblemPressure}. From which it can be observed that at the initial stage of the calculation, the incident shock wave reflects off the triangular edge, and the shock wave pattern at the wall is similar to that of a double Mach reflection. After the shock wave passes the rear corner, the shock wave diffracts, while the shock waves on the upper and lower sides converge, creating complex flow phenomena. Comparing the results of the first-order and second-order algorithms, we can see that the second-order algorithm captures more details in the flow field around the triangular corner. Subsequently, in Fig. \ref{Schileren_Comparison}, the Schlieren diagram obtained by the present DGM with N=1 is shown, in which the experimental results are given as a reference. The figure shows that the computed shock wave shape and evolution process agree well with the experiments. The triple point trajectory of the shock wave and the vortex centre trajectory are shown in Fig. \ref{SchardinProblem_trajectory}. The Mach distribution along the x-axis at time $t=172\mathrm{\mu s}$ is shown in Fig. \ref{SchardinProblem_Mach_distribution}. From these figures, it can be concluded that the results of this paper are consistent with experimental and numerical simulation results from the literature, validating the correctness of the proposed algorithm.

\subsection{Blunt-cylinder}
\begin{figure}[htpb]
\centering
\includegraphics[width=6.6in]{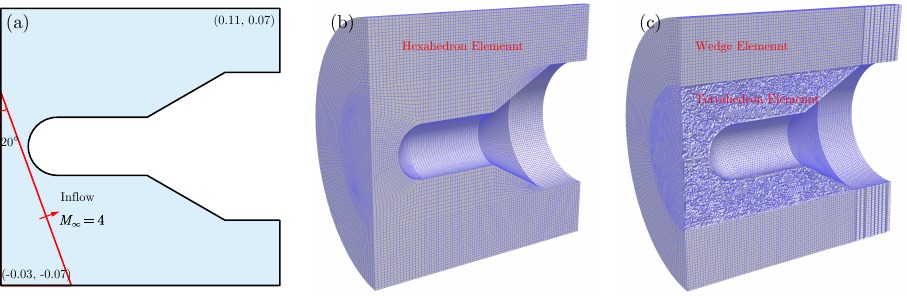}
\caption{Problem description and mesh discretization, (a) dimension of the computational domain; (b) hexahedron meshes; (c) mixed tetrahedral and wedge meshes}
\label{BluntCylinderSchematicDiagram}
\end{figure}

This problem has been studied experimentally and numerically in \cite{houtman1995experimental}. The coordinate system and the dimensions of the model are shown in Fig. \ref{BluntCylinderSchematicDiagram} (a). In the present work, this problem is studied by two kinds of mesh discretization, and the mesh is shown in Fig. \ref{BluntCylinderSchematicDiagram} (b) and (c). For the hexahedron discretization in Fig. \ref{BluntCylinderSchematicDiagram} (b), the mesh size is 1.5mm, and the total element number is 514976. Moreover, for the hybrid discretization in Fig. \ref{BluntCylinderSchematicDiagram} (c), the minimum mesh size is 1.5mm, and the total element number is 2943362.

The boundary conditions are set as follows: the left boundary is set as an inflow boundary condition with a Mach number of 4.0 and an angle of attack of 20°, the density is $\rho = 1.4$, pressure is $p=1.0$, the right boundary is set as an outflow boundary condition, the circular surface of the cylinder is set as a slip wall boundary condition. The simulation was performed up to $t=0.1$.The flow was allowed to develop until it reached a steady state. 

\begin{figure}[htpb]
\centering
\includegraphics[width=6.6in]{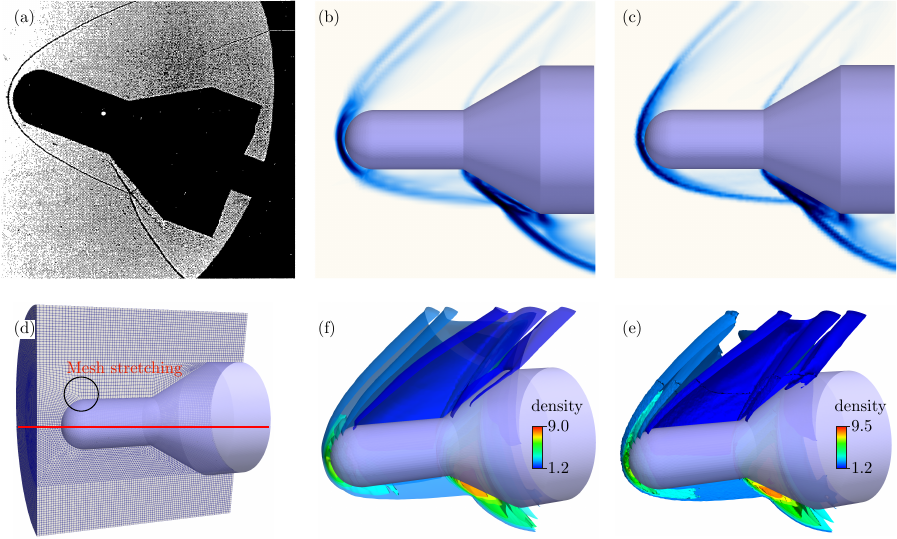}
\caption{The experimental and numerical results are presented as follows: (a)-(c) show the Schlieren images obtained from the experiment \cite{houtman1995experimental}, as well as the numerical results using hexahedral and mixed meshes; (d) presents a comparison of the meshes; (e)-(f) display the density contour surfaces obtained using hexahedral and mixed meshes.}
\label{fig-blunt-result}
\end{figure}

The results are presented in Fig. \ref{fig-blunt-result}. It is evident that the shock-shock interaction in the windward region, which affects the re-entry body, is effectively captured by the present method. A comparison between the Schlieren results in Fig. \ref{fig-blunt-result}(b) and (c) reveals that the computed shock location in the flow field aligns well with the experimental observations. However, when using hexahedral elements to discretize the flow field, some elements are stretched, as shown in Fig. \ref{fig-blunt-result}(d), leading to minor distortions in the calculated Schlieren images. In contrast, when tetrahedral and wedge elements are employed, the mesh size remains relatively uniform, resulting in more accurate Schlieren results. Since the tetrahedral discretization requires a higher number of elements, it captures the shock interface with greater sharpness. The density contour surfaces generated by the two types of meshes are shown in Fig. \ref{fig-blunt-result}(e) and (f), where the contour surfaces produced by the hexahedral mesh appear smoother. Overall, the numerical results demonstrate a good agreement with the experimental data, confirming the robustness and accuracy of the present solver in addressing the three-dimensional complex flow problem.

\section{Conclusion}
In the present work, we construct orthogonal polynomials for two-dimensional and three-dimensional isoparametric elements based on Jacobi polynomials. The modal expansion technique achieves a high-order spatial expansion of physical quantities within the element, and the discrete governing equations for the Discontinuous Galerkin Method (DGM) are derived. Building on this, we introduce the algorithm for projecting the fluid flux onto the element boundary normals, and the construction scheme for boundary numerical flux on unstructured grids is developed. Finally, an improved discontinuity-capturing technique is employed to identify discontinuous elements; the gradient of the physical quantities in these elements is limited.

Subsequently, the accuracy and convergence rate of the proposed DGM solver were evaluated using several benchmark cases. The results indicate that the adoption of high-order orthogonal polynomial expansions significantly improves convergence rates. Moreover, the computational efficiency of the algorithm is analyzed by comparing the CPU time required for different polynomial orders. The findings demonstrate that high-order methods achieve comparable error levels with substantially coarser grids, thereby challenging the conventional perception that high-order methods are inherently computationally expensive. This efficiency stems from their ability to achieve higher accuracy with fewer degrees of freedom. Finally, an arbitrary high-order DG solver was developed based on the framework established in this study. Its effectiveness and accuracy were validated through simulations of various compressible flow cases, underscoring its potential for high-fidelity computational fluid dynamics applications.

\section*{Acknowledgements}
The authors thank the National Natural Science Foundation of China (51909042) for their support.


\bibliographystyle{elsarticle-num}
\bibliography{DGEuler_references}



\end{document}